\pdfoutput=1
\documentclass{PoS}
\usepackage[utf8]{inputenc}
\usepackage{microtype}
\usepackage{amsmath}
\usepackage{booktabs}
\usepackage{physics}
\usepackage{siunitx}
\usepackage{wrapfig}
\usepackage{pgf}
\usepackage[caption=false]{subfig}

\newcommand{\fm}{\femto\metre}

\newcommand{\I}{\mathrm{i}}

\newcommand{\dmu}{\hat{\mu}}
\newcommand{\dnu}{\hat{\nu}}

\title{Non-Gaussianity of the topological charge distribution in $\mathrm{SU}(3)$ Yang--Mills theory}

\ShortTitle{Non-Gaussianity of the topological charge distribution in $\mathrm{SU}(3)$ Yang--Mills theory}

\author{\speaker{Marco Cè}\\
        Scuola Normale Superiore, Pisa \& INFN - Sezione di Pisa\\
        E-mail: \email{marco.ce@sns.it}}

\abstract{In Yang--Mills theory, the cumulants of the naïve lattice discretization of the topological charge evolved with the Yang--Mills gradient flow coincide, in the continuum limit, with those of the universal definition. We sketch in these proceedings the main points of the proof. By implementing the gradient-flow definition in numerical simulations, we report the results of a precise computation of the second and the fourth cumulant of the $\mathrm{SU}(3)$ Yang--Mills theory topological charge distribution, in order to measure the deviation from Gaussianity. A range of high-statistics Monte Carlo simulations with different lattice volumes and spacings is used to extrapolate the results to the continuum limit with confidence by keeping finite-volume effects negligible with respect to the statistical errors. Our best result for the topological susceptibility is $t_0^2\chi=\num{6.67(7)e-4}$, while for the ratio between the fourth and the second cumulant we obtain $R=\num{0.233(45)}$.}

\FullConference{The 33rd International Symposium on Lattice Field Theory\\
		14 -18 July 2015\\
		Kobe International Conference Center, Kobe, Japan*}

\begin{document}
\section{Introduction}
The Witten--Veneziano (WV) mechanism~\cite{Witten:1979vv,Veneziano:1979ec} relates the mass of the $\eta'$ meson to the topological susceptibility in Yang--Mills (YM) theory, a quantity that can be measured on the lattice. However, the naïve lattice discretization of the topological charge density needs an unambiguous renormalization condition to be correctly defined. When considering its cumulants, such as the susceptibility, short-distance singularities contribute also to an additive renormalization. On the contrary, if defined using Ginsparg--Wilson (GW) fermions~\cite{Neuberger:1997fp,Hasenfratz:1998ri,Luscher:1998pqa}, the topological charge and its cumulants have finite and unambiguous continuum limits without renormalization, which satisfy the anomalous chiral Ward identities behind the WV mechanism~\cite{Giusti:2001xh,Giusti:2004qd,Luscher:2004fu}. 

The recent proposal of evolving the naïve discretization with the YM gradient flow~\cite{Luscher:2010iy} is an alternative definition with a finite and unambiguous continuum limit~\cite{Luscher:2010iy,Luscher:2011bx}, and is particularly appealing because its numerical evaluation is significantly cheaper than the definition using Neuberger's GW satisfying fermions.

In these proceedings we discuss our recent results~\cite{Ce:2015qha}. In Section~\ref{sec:definition_equivalence} we show that in the YM theory the cumulants of the topological charge with the naïve discretization at positive flow time coincide, in the continuum limit, with those of the universal definition. In the following sections, by implementing the gradient-flow definition we compute the topological susceptibility with a precision $5$ times better than the reference computation with Neuberger's definition~\cite{DelDebbio:2004ns} and the ratio $R$ between the fourth and the second cumulant with all systematics negligible with respect to the statistical error.

\section{Cumulants of the topological charge on the lattice}
\label{sec:definition_equivalence}
In the continuum, given the topological nature of the topological charge
\begin{equation}
\label{eq:top_charge_def}
  Q = \int \dd[4]{x} q(x)\;, \qquad q(x) = \frac{1}{32\pi^2} \epsilon_{\mu\nu\rho\sigma} \tr{F_{\mu\nu}F_{\rho\sigma}}\;,
\end{equation}
it's easy to see that $Q$ is invariant under continuous deformation of the fields induced by the YM gradient flow. Indeed, denoting with a superscript $t$ the flow-time dependence\footnote{If not explicitly indicated, we assume $t>0$.},
\begin{equation}
\label{eq:q_dflow}
  \partial_t q^t = \partial_\rho w_\rho^t\;, \qquad w_\rho^t = \frac{1}{8\pi^2} \epsilon_{\mu\nu\rho\sigma} \tr{F^t_{\mu\nu}D_\alpha F^t_{\alpha\sigma}}\;,
\end{equation}
where $w_\rho^t$ is a local dimension-$5$ gauge-invariant pseudovector field, thus $\partial_t Q^t=0$.

Eq.~\eqref{eq:q_dflow} implies, for any finite (multi)local field $O(y)$ at a physical distance from $x$,
\begin{equation}
  \ev{q^t(x) O(y)} = \ev{q^{t=0}(x) O(y)} + \partial_\rho \int_0^t \dd{t'} \ev{w_\rho^{t'}(x) O(y)}\;, \qquad x\neq y\;.
\end{equation}
The $t=0$ limit of this equation is taken as the definition of the renormalized $q(x)$. For $x\neq y$, there are no local fields of $d<5$ that $w_\rho^t$ can mix with, thus
\begin{equation}
  \ev{q^{t=0}(x) O(y)} = \lim_{t\to0} \ev{q^t(x) O(y)}.
\end{equation}

We are interested in cumulants of the topological charge
\begin{equation}
  C_n^t = \int \dd[4]{z_1}\dots\dd[4]{z_{2n-1}} \ev{q^t(0)q^t(z_1)\dots q^t(z_{2n-1})}\;.
\end{equation}
To study the short-distance singularities at $t=0$, we supplement the theory with degenerate valence quarks and substitute a single $q^{t=0}(z_i)$ with a chain of renormalized (pseudo)scalar densities~\cite{Luscher:2004fu}
\begin{equation}
\label{eq:density_chain}
  \int \dd[4]{z} q^{t=0}(z) \to -m^5 \int \left( \prod \dd[4]{x_i} \right) \ev{P_{51}(x_1)S_{12}(x_2)S_{23}(x_3)S_{34}(x_4)S_{45}(x_5)}_\text{F}\;.
\end{equation}
When one of the $x_j$ is close to $0$, according to the operator product expansion
\begin{equation}
  q^{t=0}(0)S_{ij}(x) \xrightarrow{\abs{x}\to 0} c(x) P_{ij}(0) + \dots\;,
\end{equation}
where $c(x)$ is a function that diverges as $\abs{x}^{-4}$ for $\abs{x}\to 0$ and the dots indicate subleading contributions. The Wilson coefficient of the leading short-distance singularity can be computed in perturbation theory. From Eq.~\eqref{eq:q_dflow} we have
\begin{equation}
  \ev{q^t(0) S_{ij}(x) O(y)} = \ev{q^{t=0}(0) S_{ij}(x) O(y)} + \partial_\rho \int_0^t \dd{t'} \ev{w_\rho^{t'}(0) S_{ij}(x) O(y)}\;,
\end{equation}
where again $O(y)$ is any finite (multi)local field at a physical distance from $0$ and $x$. When $t>0$, the l.h.s. has no singularities for $\abs{x}\to 0$, thus singularities must cancel in the r.h.s. and $c(x)$ must be of the form
\begin{equation}
\label{eq:wilson_coeff}
  c(x) = \partial_\rho u_\rho(x)
\end{equation}
which does not contribute when integrated over spacetime, and leads to $\partial_t C_n^t=0$ for every $t\geq 0$.

\subsection{Ginsparg--Wilson discretization of cumulants}
On the lattice, the gauge field evolved with the YM gradient flow is smooth on the scale of the cut-off. Gauge-invariant local fields constructed with the gauge field at positive flow time are finite as they stand, with the universality class determined only by their asymptotic behaviour in the classical continuum limit~\cite{Luscher:2010iy,Luscher:2011bx}. Short-distance singularities cannot arise because of the exponential damping at high frequencies enforced with the flow. Therefore, to show that cumulants of the topological charge at $t>0$ satisfy the proper singlet chiral Ward identities it is sufficient to work with a particular discretization of $q$. The GW definition $q_\text{GW}(x)$~\cite{Neuberger:1997fp,Hasenfratz:1998ri,Luscher:1998pqa} has a privileged rôle\footnote{A concrete example is the definition with Neuberger's fermions~\cite{Neuberger:1997fp}: $a^4q_\text{N}(x)=-(\flatfrac{\bar{a}}{2})\tr\gamma_5D_\text{N}(x,x)$.} since the bare lattice cumulants with the GW lattice discretization,
\begin{equation}
  C_{\text{GW},n}^t = a^{4(2n-1)} \sum_{x_i} \ev{q_\text{GW}^t(0)q_\text{GW}^t(x_1)\dots q_\text{GW}^t(x_{2n-1})}\;,
\end{equation}
are finite and satisfy the anomalous chiral Ward identity for $t=0$~\cite{Giusti:2001xh,Giusti:2004qd,Luscher:2004fu}.

This is not obvious at $t>0$. To prove it, first consider the susceptibility $C_{\text{GW},1}^{t=0}$ and apply the flow to one of the two densities. While no short-distance singularities are present and $q_\text{GW}^t$ is finite as it stands, $q_\text{GW}^{t=0}$ can have a renormalization constant $Z_q$\footnote{$Z_q$ is at most logarithmically divergent, since there are no other pseudoscalar gauge-invariant operators with $d\leq 4$.}. Using the lattice equivalent of Eq.~\eqref{eq:density_chain}, it's possible to show that $q_\text{GW}^{t=0}$ with  $Z_q=1$ converges to the correct continuum limit with rate $a^2$.

Then, consider again the susceptibility $C_{\text{GW},1}^{t=0}$ and replace one of the two $q_\text{GW}^{t=0}$ with its lattice density-chain expression. The discussion in the continuum case applies here and Eq.~\eqref{eq:wilson_coeff} guarantees that there are no short-distance singularities. This result and $Z_q=1$ imply that
\begin{equation}
  \lim_{t\to 0}\lim_{a\to 0} a^4\sum_x \ev{q_\text{GW}^t(0) q_\text{GW}^{t=0}(x)} = \lim_{a\to 0} a^4\sum_x \ev{q_\text{GW}^{t=0}(0) q_\text{GW}^{t=0}(x)}\;.
\end{equation}
By replacing the $t=0$ density on the l.h.s. with the evolved one, no further singularities arise. The same argument applies to $C_{\text{GW},n}^t$ replacing $2n-1$ charges with their density-chain expressions, and we arrive to the result
\begin{equation}
\label{eq:GW_equivalence}
  \lim_{t\to 0} \lim_{a\to 0} C_{\text{GW},n}^t = \lim_{a\to 0} C_{\text{GW},n}^{t=0}\;.
\end{equation}

\subsection{Universality at positive flow time}
The naïve discretization of the topological charge density $q_\text{L}(x)$ is the lattice version of Eq.~\eqref{eq:top_charge_def} in which a discretized field strength tensor is used
\begin{equation}
  F_{\mu\nu}^a(x) = -\frac{\I}{4a^2} \tr{[Q_{\mu\nu}(x)-Q_{\nu\mu}(x)]T^a}\;,
\end{equation}
with
\begin{equation}
  \begin{split}
    Q_{\mu\nu}(x) = &\, U_\mu(x)               U_\nu(x+a\dmu)               U^\dagger_\mu(x+a\dnu)     U^\dagger_\nu(x) \\
                  + &\, U_\nu(x)               U^\dagger_\mu(x-a\dmu+a\dnu) U^\dagger_\nu(x-a\dmu)     U_\mu(x-a\dmu)   \\
                  + &\, U^\dagger_\mu(x-a\dmu) U^\dagger_\nu(x-a\dmu-a\dnu) U_\mu(x-a\dmu-a\dnu)       U_\nu(x-a\dnu)   \\
                  + &\, U^\dagger_\nu(x-a\dnu) U_\mu(x-a\dnu)               U_\nu(x+a\dmu-a\dnu)       U^\dagger_\mu(x)\;.
  \end{split}
\end{equation}
At $t=0$, $q_\text{L}^{t=0}(x)$ require a multiplicative renormalization constant~\cite{Alles:1996nm}. The corresponding cumulants $Q_\text{L}^{t=0}$ have additional ultraviolet power-divergent singularities and they do not have a well defined continuum limit. Conversely, at positive flow time the density $q_\text{L}^t(x)$ shares with $q_\text{GW}^t(x)$ the same asymptotic behaviour in the classical continuum limit. The universality at positive flow time together with Eq.~\eqref{eq:GW_equivalence} imply that the topological charge cumulants discretized with the naïve definition at positive flow time
\begin{equation}
  \lim_{t\to 0} \lim_{a\to 0} C_{\text{L},n}^t = \lim_{a\to 0} C_{\text{GW},n}^{t=0}
\end{equation}
and satisfies the anomalous chiral Ward identities. It is interesting to note, however, that at fixed lattice spacing there can be differences\footnote{For instance, the topological susceptibility with the naïve definition at $t>0$ is not guaranteed to go to zero in the chiral limit at finite lattice spacing in the presence of fermions.}.

\section{Numerical setup}
\begin{table}[tb]
  \centering
  {\small\begin{tabular}{cS[table-format=1.2]S[table-format=2]S[table-format=1.1]S[table-format=1.3]S[table-format=7]S[table-format=3]S[table-format=1.5(2)]S[table-format=1.3(2)]S[table-format=1.3(2)]}
    \toprule
    Lattice & {$\beta$} & {$\flatfrac{L}{a}$} & {$L$\,[\si{\fm}]} & {$a$\,[\si{\fm}]} & {$N_\text{conf}$} & {$N_\text{it}$} & {$\flatfrac{t_0}{a^2}$} & {$L^4\chi$} & {$R$} \\
    \midrule
    $A_1$ &    5.96   &    10   &  1.0  &  0.102  &      36000       &  30 &2.995(4)     & 0.701(6)  & 0.39(3)   \\
    $B_1$ &           &    12   &  1.2  &         &     144000       &     & 2.7984(9)   & 1.617(6)  & 0.187(24) \\
    $C_1$ &           &    13   &  1.3  &         &     280000       &     & 2.7908(5)   & 2.244(6)  & 0.177(23) \\
    $D_1$ &           &    14   &  1.4  &         &     505000       &     & 2.7889(3)   & 3.028(6)  & 0.209(23) \\
    $E_1$ &           &    15   &  1.5  &         &     880000       &     & 2.78892(23) & 3.982(6)  & 0.202(23) \\
    $F_1$ &           &    16   &  1.6  &         &    1440000       &     & 2.78867(16) & 5.167(6)  & 0.157(22) \\
    \midrule
    $B_2$ &    6.05   &    14   &  1.2  &  0.087  &     144000       &  60 & 3.7960(12)  & 1.699(7)  & 0.24(3)   \\
    $D_2$ &           &    17   &  1.5  &         &     144000       &     & 3.7825(8)   & 3.686(14) & 0.22(5)   \\
    \midrule
    $B_3$ &    6.13   &    16   &  1.2  &  0.077  &     144000       &  90 & 4.8855(15)  & 1.750(7)  & 0.22(3)   \\
    $D_3$ &           &    19   &  1.5  &         &     144000       &     & 4.8722(11)  & 3.523(13) & 0.16(5)   \\
    \midrule
    $B_4$ &    6.21   &    18   &  1.2  &  0.068  &     144000       & 250 & 6.2191(20)  & 1.741(7)  & 0.20(3)   \\
    $D_4$ &           &    21   &  1.4  &         &     144000       &     & 6.1957(14)  & 3.266(12) & 0.21(5)   \\
    \bottomrule
  \end{tabular}}
  \caption{Ensembles and statistics used in this study together with measured observables. For each lattice we give the label, $\beta=\flatfrac{6}{g_0^2}$, the spatial extent of the lattice, the lattice spacing, the number $N_\text{conf}$ of independent configurations generated, the number of sweeps $N_\text{it}$ required to space them, the measured reference scale $\flatfrac{t_0}{a^2}$, the susceptibility $L^4\chi$ and the ratio $R$.}
  \label{tab:simulation_details}
\end{table}

To compute the topological susceptibility $\chi^t\equiv C_{\text{L},1}^t$ and the fourth cumulant $C_{\text{L},2}^t$, we discretize $\mathrm{SU}(3)$ YM theory using the standard Wilson plaquette action on a finite lattice of size $\flatfrac{L}{a}$ in all four spacetime directions and periodic boundary conditions. The Monte Carlo updates the gauge field implementing the Cabibbo--Marinari scheme, with a heat bath sweep of all lattice links followed by $\flatfrac{L}{(2a)}$ over-relaxation sweeps. We simulate the lattices listed with their parameters in Table~\ref{tab:simulation_details}.

We use the $\{B_1\dots B_4\}$ series, with physical volume of $\sim(\SI{1.2}{\fm})^4$, to extrapolate the continuum limit of the fourth cumulant. The choice of a very small volume is forced by the fact that statistical error on $C_{\text{L},2}$ are $\order{V}$. Given the higher statistical accuracy, we extrapolate the continuum limit of the susceptibility from the $\{D_1\dots D_4\}$ series, with physical volume of $\sim(\SI{1.4}{\fm})^4$. The other lattices are used to check that the finite-volume systematics is under control and negligible with respect to the final statistical error.

High-precision scale setting is obtained measuring the energy density $E^t$ at positive flow time to compute the reference flow time $t_0$~\cite{Luscher:2010iy} as the solution of
\begin{equation}
\label{eq:t0_def}
  \left. t^2 \ev{E^t} \right|_{t=t_0} = 0.3\;, \qquad E^t = \frac{1}{4V} \sum_x F_{\mu\nu}^{a,t}(x) F_{\mu\nu}^{a,t}(x)\;.
\end{equation}

The YM gradient flow equation is integrated numerically implementing a fourth order structure-preserving Runge--Kutta--Munthe-Kaas method\footnote{See Appendix~B of Ref.~\cite{Ce:2015qha} for the details.}. On each lattice, the gauge field is evolved approximatively up to $\sim 1.2t_0$. Two different RK step size are used to check the systematics from the numerical integration and a smaller step size is used on per-configuration basis to have negligible systematic errors with respect to the statistical error.

Dedicated simulations are performed to measure the integrated autocorrelation time $\tau_\text{int}$ of $Q$, its moments and $t_0$. In the range of $\beta$ considered, $Q$ has the largest $\tau_\text{int}$ which increases rapidly toward the continuum limit. The number of Cabibbo--Marinari updates between measurements in the main runs, $N_\text{it}$ in Table~\ref{tab:simulation_details}, is chosen to have statistically independent measurements of $Q$.

Observables are measured at $t=t_0$ and given in terms of the two dimensionless quantities $t_0^2\chi\equiv t_0^2 C_{\text{L},1}^{t_0}$ and ratio $R\equiv\flatfrac{C_{\text{L},2}^{t_0}}{C_{\text{L},1}^{t_0}}$ between the fourth and second cumulant. Bare lattice results are reported in Table~\ref{tab:simulation_details}.

\section{Results}
\begin{figure}[t]
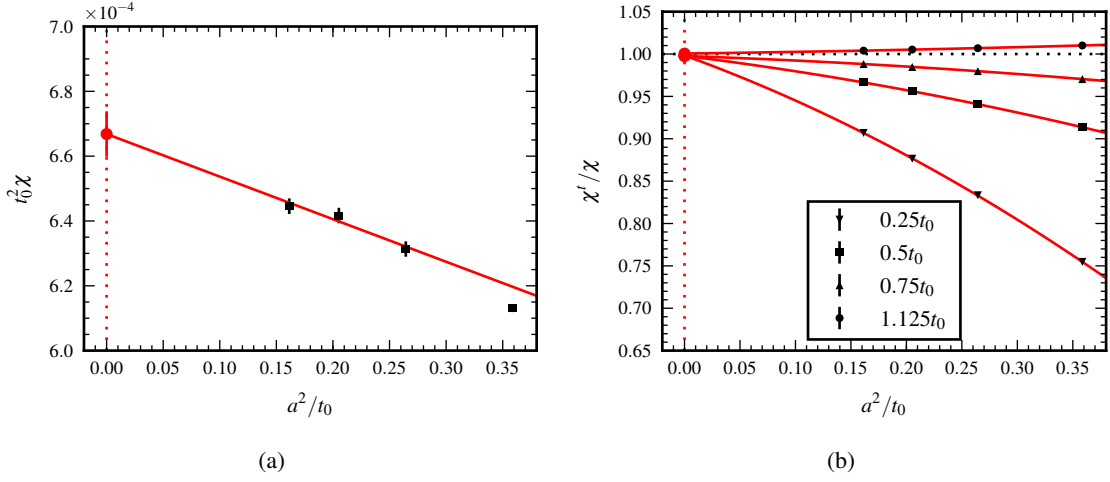

  \subfloat[\label{fig:continuum_limit_chi}]{\input{graphics/continuum_limit_chi.pgf}}
  \subfloat[\label{fig:continuum_limit_chi_ratio}]{\input{graphics/continuum_limit_chi_ratio.pgf}}
  \caption{(a) Topological susceptibility $t_0^2\chi$ with extrapolation to the continuum limit. (b) The ratio $\flatfrac{\chi^t}{\chi}$ (errors are smaller than symbols) for several values of $t$, and its extrapolation to the continuum limit.}
  \label{fig:continuum_limit}
\end{figure}

\begin{wrapfigure}{R}{0.5\columnwidth}
  \vspace{-1.7cm}
  \input{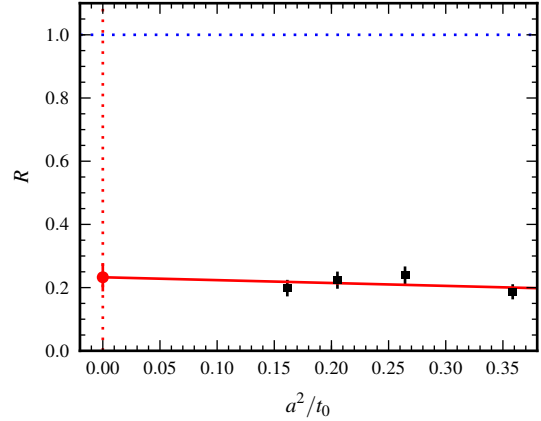}
  \caption{The ratio $R$ with continuum extrapolation. In blue the dilute instanton gas model prediction $R=1$.}
  \label{fig:continuum_limit_R}
\end{wrapfigure}

At the coarsest lattice spacing, no systematic volume effects are observed in the value of $t_0^2\chi$ starting from $V\sim(\SI{1.4}{\fm})^4$. Thus, the continuum value of $t_0^2\chi$ is extrapolated in Figure~\ref{fig:continuum_limit_chi} from the values measured on the lattices $\{D_1,\dots,D_4\}$, fitting with $\order{a^2}$ discretization errors according to Symanzik effective theory. The best fit is obtained with a linear fit restricted to the three finest lattices and gives
\begin{equation}
\label{eq:continuum_limit_chi}
  t_0^2 \chi = \SI{6.67(7)e-4}\;.
\end{equation}
A linear fit including the coarsest lattice and a quadratic fit of all points give compatible results.

The cumulants of the topological charge are expected to be $t$-independent in the continuum limit. In Figure~\ref{fig:continuum_limit_chi_ratio} we show the topological susceptibility computed at various $t$ normalized to its value at $t_0$. The data points have statistical errors in the range $0.1$--$1$ per mille, due to the correlation between numerators and denominators. $t$-dependent discretization effects are fitted with a quadratic function in $\flatfrac{a^2}{t_0}$, obtaining for each $t$ an intercept compatible with $1$ within statistical errors in the range $0.5$--$5$ per mille. This represent a high-precision test of universality in the continuum limit of the gradient-flow discretization of the topological charge.

The continuum value of $R$ is extrapolated in Figure~\ref{fig:continuum_limit_R} from the results from the $V\sim(\SI{1.2}{\fm})^4$ lattices $\{B_1,\dots,B_4\}$ without visible finite-volume systematics. The result of a linear fit in $a^2$ is
\begin{equation}
\label{eq:continuum_limit_R}
  R = \num{0.233(45)}\;,
\end{equation}
with a compatible result obtained from a constant fit.

\section{Conclusions}
In Ref.~\cite{Ce:2015qha} we prove that, in the continuum limit, the cumulants of the topological charge defined by the Yang--Mills gradient flow coincide with those of the universal definition appearing in the chiral Ward identities. We use this definition to study the topological charge distribution with an unprecedented precision. Our result in Eq.~\eqref{eq:continuum_limit_R} is the first one for the ratio $R$ with systematic and statistical errors under control. The value in Eq.~\eqref{eq:continuum_limit_R} agrees with the result of Ref.~\cite{Giusti:2007tu}, it is inconsistent with the dilute instanton gas model prediction of $R=1$, and it is compatible with the large-$N_c$ prediction of being of order $\flatfrac{1}{N_c^2}$. We also measure the topological susceptibility with unprecedented precision. The value in Eq.~\eqref{eq:continuum_limit_chi} is compatible with other lattice results \cite{DelDebbio:2004ns,Luscher:2010ik}. As a by-product, we obtain a sub-percent numerical test of universality of the gradient-flow definition.

\bibliographystyle{JHEP}
\bibliography{biblio.bib}

\providecommand{\href}[2]{#2}\begingroup\raggedright\begin{thebibliography}{10}

\bibitem{Witten:1979vv}
E.~Witten, {\it Current algebra theorems for the {$U(1)$} ``{Goldstone}
  boson''},  {\em Nucl. Phys. B} {\bf 156} (1979) 269--283.

\bibitem{Veneziano:1979ec}
G.~Veneziano, {\it {$U(1)$} without instantons},  {\em Nucl. Phys. B} {\bf 159}
  (1979) 213--224.

\bibitem{Neuberger:1997fp}
H.~Neuberger, {\it Exactly massless quarks on the lattice},  {\em Phys. Lett.
  B} {\bf 417} (1998) 141--144,
  [\href{http://arxiv.org/abs/hep-lat/9707022}{{\tt hep-lat/9707022}}].

\bibitem{Hasenfratz:1998ri}
P.~Hasenfratz, V.~Laliena, and F.~Niedermayer, {\it The index theorem in {QCD}
  with a finite cut-off},  {\em Phys. Lett. B} {\bf 427} (1998) 125--131,
  [\href{http://arxiv.org/abs/hep-lat/9801021}{{\tt hep-lat/9801021}}].

\bibitem{Luscher:1998pqa}
M.~Lüscher, {\it Exact chiral symmetry on the lattice and the
  {Ginsparg--Wilson} relation},  {\em Phys. Lett. B} {\bf 428} (1998) 342--345,
  [\href{http://arxiv.org/abs/hep-lat/9802011}{{\tt hep-lat/9802011}}].

\bibitem{Giusti:2001xh}
L.~Giusti, G.~C. Rossi, M.~Testa, and G.~Veneziano, {\it The {$U_A(1)$} problem
  on the lattice with {Ginsparg-Wilson} fermions},  {\em Nucl. Phys. B} {\bf
  628} (2002) 234--252, [\href{http://arxiv.org/abs/hep-lat/0108009}{{\tt
  hep-lat/0108009}}].

\bibitem{Giusti:2004qd}
L.~Giusti, G.~Rossi, and M.~Testa, {\it Topological susceptibility in full
  {QCD} with {Ginsparg--Wilson} fermions},  {\em Phys. Lett. B} {\bf 587}
  (2004) 157--166, [\href{http://arxiv.org/abs/hep-lat/0402027}{{\tt
  hep-lat/0402027}}].

\bibitem{Luscher:2004fu}
M.~Lüscher, {\it Topological effects in {QCD} and the problem of
  short-distance singularities},  {\em Phys. Lett. B} {\bf 593} (2004)
  296--301, [\href{http://arxiv.org/abs/hep-th/0404034}{{\tt hep-th/0404034}}].

\bibitem{Luscher:2010iy}
M.~Lüscher, {\it Properties and uses of the {Wilson} flow in lattice {QCD}},
  {\em JHEP} {\bf 1008} (2010) 071, [\href{http://arxiv.org/abs/1006.4518}{{\tt
  arXiv:1006.4518}}].

\bibitem{Luscher:2011bx}
M.~Lüscher and P.~Weisz, {\it Perturbative analysis of the gradient flow in
  non-abelian gauge theories},  {\em JHEP} {\bf 1102} (2011) 051,
  [\href{http://arxiv.org/abs/1101.0963}{{\tt arXiv:1101.0963}}].

\bibitem{Ce:2015qha}
M.~Cè, C.~Consonni, G.~P. Engel, and L.~Giusti, {\it Non-{Gauss}ianities in
  the topological charge distribution of the {$SU(3)$} {Yang-Mills} theory},
  {\em Phys. Rev. D} {\bf 92} (2015)
  [\href{http://arxiv.org/abs/1506.06052}{{\tt arXiv:1506.06052}}].

\bibitem{DelDebbio:2004ns}
L.~Del~Debbio, L.~Giusti, and C.~Pica, {\it Topological susceptibility in the
  {$SU(3)$} gauge theory},  {\em Phys. Rev. Lett.} {\bf 94} (2005) 032003,
  [\href{http://arxiv.org/abs/hep-th/0407052}{{\tt hep-th/0407052}}].

\bibitem{Alles:1996nm}
B.~Allés, M.~D'Elia, and A.~D. Giacomo, {\it Topological susceptibility at
  zero and finite t in {$\text{SU}(3)$} {Yang-Mills} theory},  {\em Nucl. Phys.
  B} {\bf 494} (1997), no.~1-2 281--292,
  [\href{http://arxiv.org/abs/hep-lat/9605013}{{\tt hep-lat/9605013}}].

\bibitem{Giusti:2007tu}
L.~Giusti, S.~Petrarca, and B.~Taglienti, {\it {$\theta$} dependence of the
  vacuum energy in {$SU(3)$} gauge theory from the lattice},  {\em Phys. Rev.
  D} {\bf 76} (2007) 094510, [\href{http://arxiv.org/abs/0705.2352}{{\tt
  arXiv:0705.2352}}].

\bibitem{Luscher:2010ik}
M.~Lüscher and F.~Palombi, {\it Universality of the topological susceptibility
  in the {SU(3)} gauge theory},  {\em JHEP} {\bf 1009} (2010) 110,
  [\href{http://arxiv.org/abs/1008.0732}{{\tt arXiv:1008.0732}}].

\end{thebibliography}\endgroup

\end{document}